\documentclass[journal]{IEEEtran}
\usepackage{cite}
\usepackage{amsmath,amssymb,amsfonts}
\usepackage{algorithmic}
\usepackage{graphicx}
\usepackage{textcomp}
\usepackage[inline]{enumitem}
\usepackage{balance}
\usepackage{booktabs}
\usepackage{multirow}

\usepackage{threeparttable}
\makeatletter
\let\MYcaption\@makecaption
\makeatother
\usepackage[font=footnotesize]{subcaption}
\makeatletter
\let\@makecaption\MYcaption
\makeatother
\usepackage{caption}
\newcommand{\dd}        {\mathrm{d}}
\newcommand{\ee}        {\mathrm{e}}
\newcommand{\jj}        {\mathrm{j}}
\def\BibTeX{{\rm B\kern-.05em{\sc i\kern-.025em b}\kern-.08em
    T\kern-.1667em\lower.7ex\hbox{E}\kern-.125emX}}
\begin{document}
%\history{Date of publication xxxx 00, 0000, date of current version xxxx 00, 0000.}
%\doi{TBA}

\title{Accurate Radar-Based Heartbeat Measurement Using Higher Harmonic Components}

\author{Itsuki~Iwata, Kimitaka~Sumi, Yuji~Tanaka,~\IEEEmembership{Member, IEEE,} and Takuya~Sakamoto,~\IEEEmembership{Senior Member, IEEE,}
\thanks{I.~Iwata, K.~Sumi, and T.~Sakamoto are with the Department of Electrical Engineering, Graduate School of Engineering, Kyoto University, Kyoto 615-8510, Japan.}
\thanks{Y.~Tanaka is with Graduate School of Engineering, Nagoya Institute of Technology, Nagoya 466-8555, Japan.}}
%\markboth{}%
%{Sakamoto \emph{et al.}: Radar-Based Respiratory Measurement of a Rhesus Monkey by Suppressing Nonperiodic Body Motion Components}

%\address[1]{Department of Electrical Engineering,
%    Graduate School of Engineering, Kyoto University, Kyoto 615-8510, Japan}
%\address[2]{Graduate School of Engineering, Nagoya Institute of Technology, Nagoya 466-8555, Japan}
    
%\tfootnote{This work was supported in part by the SECOM Science and Technology Foundation, in part by the Japan Science and Technology Agency under Grants JPMJMI22J2 and JPMJMS2296, and in part by the Japan Society for the Promotion of Science KAKENHI under Grant 19H02155, Grant 21H03427, Grant 23H01420, Grant 23K19119, Grant 23K26115, and Grant 24K17286.}

\markboth{}
{Iwata \emph{et al.}: Accurate Radar-based Heartbeat Measurement Using Higher Harmonic Components}

%\corresp{Corresponding author: T. Sakamoto (e-mail: sakamoto.takuya.8n@kyoto-u.ac.jp).}

\maketitle

\begin{abstract}
This study proposes a radar-based heartbeat measurement method that uses the absolute value of the second derivative of the complex radar signal, rather than its phase, and the variational mode extraction method, which is a type of mode decomposition algorithm. We show that the proposed second-derivative-based approach can amplify the heartbeat component in radar signals effectively and also confirm that use of the variational mode extraction method represents an efficient way to emphasize the heartbeat component amplified via the second-derivative-based approach. We demonstrate estimation of the heart interbeat intervals using the proposed approach in combination with the topology method, which is an accurate interbeat interval estimation method. The performance of the proposed method is evaluated quantitatively using data obtained from eleven participants that were measured using a millimeter-wave radar system. When compared with conventional methods based on the phase of the complex radar signal, our proposed method can achieve higher accuracy when estimating the heart interbeat intervals; the correlation coefficient for the proposed method was increased by 0.20 and the root-mean-square error decreased by 23\%.
\end{abstract}

\IEEEpeerreviewmaketitle
%\titlepgskip=-21pt

%\maketitle

\section{Introduction}
\label{section.1}
\IEEEPARstart{R}{adar}-based technology for measurement of physiological signals is expected to change the entire concept of health management because this technology can enable continuous monitoring of the health and mental status of patients that are visible within a scene. Currently, physiological measurements for healthcare are based on the use of contact-type sensors, including electrocardiogram sensors with electrodes, pulse oximeter devices that use optical sensors, and blood pressure monitors that require a cuff to be wrapped around the patient’s arm. Use of these contact-type sensors for continuous long-term measurements can cause discomfort for the user or impose restrictions on their activities \cite{https://doi.org/10.1109/ACCESS.2019.2921240, https://doi.org/10.1109/ACCESS.2023.3302918}, but these problems can be avoided by introducing radar-based noncontact measurement technology \cite{https://doi.org/10.1109/TIM.2021.3129498}.

Radar-based measurements of physiological signals use radar echoes that are reflected from the surface of the human body and modulated by the body’s displacements. The heartbeat causes a semi-periodic displacement that can be observed over the entire human body \cite{https://doi.org/10.1109/ACCESS.2022.3211527} and the displacement due to the heartbeat can be expressed as a summation of the heart motion and the volumetric changes in the pulse waves that propagate through the arteries. 
In principle, by estimating the power spectrum of these displacements, we can detect the respiratory and heartbeat components within the frequency domain. Because these respiratory and heartbeat components are semi-periodic but are not sinusoidal, they also contain higher-order harmonics in addition to their fundamental frequencies \cite{https://doi.org/10.1109/TIM.2023.3267348, https://doi.org/10.1109/JSEN.2023.3250500}.

A typical respiratory displacement has a fundamental frequency of between 0.1 and 0.3 Hz and a displacement magnitude of between 4 and 12 mm; a typical heartbeat displacement has a fundamental frequency of between 1 and 1.7 Hz, with a displacement magnitude of between 0.2 and 0.5 mm \cite{https://doi.org/10.1002/mop.24877, https://doi.org/10.1016/j.bspc.2014.03.004, https://doi.org/10.1016/0002-9149(93)91153-9}. In the frequency domain, it is often difficult to identify the fundamental frequency of the heartbeat because this fundamental frequency component can be masked by the higher-order harmonic components of the respiration.
Therefore, a sophisticated approach is required to estimate the heartbeat component within the frequency domain, particularly when it is necessary to estimate the heart rate variability (HRV) or the heart’s interbeat intervals (IBIs) \cite{https://doi.org/10.1109/ACCESS.2022.3211527, https://doi.org/10.1109/JSEN.2023.3250500, https://doi.org/10.1109/JSEN.2019.2950635, https://doi.org/10.1109/TAES.2019.2917489, https://doi.org/10.1109/JERM.2018.2879452}. For example, Yamamoto et al. \cite{https://doi.org/10.1162/neco.1997.9.8.1735, https://doi.org/10.1109/ACCESS.2020.3006107} extracted the heartbeat component signal from radar data using a bandpass filter and a convolutional long short-term memory (LSTM) algorithm to estimate the IBIs. In addition, Wang et al. \cite{https://doi.org/10.1109/JIOT.2021.3075167, https://doi.org/10.1109/TSP.2013.2288675} extracted the heartbeat component signal from radar data using variational mode decomposition (VMD).
Petrovi\'{c} et al. \cite{https://doi.org/10.1109/ACCESS.2019.2921240} applied a bandpass filter to the complex radar signal using filter parameters that were set beforehand using a roughly estimated heart rate ; this was followed by application of a filter bank with narrow band-pass filters and resulted in IBI estimation using the zero-crossing points. 

Although these methods estimate the heartbeat parameters by focusing on the fundamental heartbeat frequency, their estimation accuracy is dependent on the intensity of the heartbeat’s fundamental frequency component, which can be masked easily by the respiratory components.
To overcome this issue, some existing studies have focused on the heartbeat’s harmonic frequency components rather than the fundamental frequency \cite{https://doi.org/10.1109/ACCESS.2019.2921240, https://doi.org/10.1109/JSEN.2019.2950635, https://doi.org/10.1109/TAES.2019.2917489, https://doi.org/10.1109/TMTT.2022.3222384}.
Rong et al. \cite{https://doi.org/10.1109/TAES.2019.2917489} reported that the second harmonic component of the heartbeat is significantly greater than the higher-order respiratory harmonics. In addition, Petrovi\'{c} et al. \cite{https://doi.org/10.1109/ACCESS.2019.2921240} showed that the heartbeat’s higher harmonics are much larger than the higher respiratory harmonics within the high frequency band. Therefore, to extract the heartbeat parameters, it is advantageous to use the higher harmonics instead of the fundamental component. Furthermore, it has been reported that use of the second harmonic component of the heartbeat is effective in suppressing the respiratory components \cite{https://doi.org/10.1109/JSEN.2019.2950635, https://doi.org/10.1109/JSEN.2022.3148003}. Iwata et al. \cite{https://doi.org/10.1109/LSENS.2023.3322287} extracted the second harmonic component of the heartbeat signal to adjust the filter parameters to ensure that they emphasized the heartbeat component.

To emphasize the heartbeat’s higher harmonics while also suppressing the respiratory components, signal processing methods such as application of a high-pass filter are necessary. To form such a high-pass filter, some studies have used the time derivative of the displacement waveform that can be estimated from the complex radar signal’s phase \cite{https://doi.org/10.1109/TIM.2023.3267348, https://doi.org/10.1109/JERM.2018.2879452, https://doi.org/10.1109/ISPA48434.2019.8966792, https://doi.org/10.1109/JERM.2023.3326562, https://doi.org/10.1109/TIM.2020.2978347, https://doi.org/10.1109/ICCECE51280.2021.9342280}.
To explain these methods, let us assume here that $s(t)$ is a complex radar signal in the time domain.
Wang et al. \cite{https://doi.org/10.1109/TIM.2023.3267348} suppressed the higher respiratory harmonics by introducing the first derivative of the radar signal phase (i.e., $(\dd/\dd t) \angle s(t)$). Kakouche et al. \cite{https://doi.org/10.1109/ISPA48434.2019.8966792} also used the first derivative of the radar signal phase ($(\dd/\dd t) \angle s(t)$) when applying the delay-and-sum method in the frequency domain. 
Yen et al. \cite{https://doi.org/10.1109/JERM.2023.3326562} also emphasized the heartbeat component by using $(\dd/\dd t) \angle s(t)$.
In contrast, Xiong et al. \cite{https://doi.org/10.1109/TIM.2020.2978347} and Ji et al. \cite{https://doi.org/10.1109/ICCECE51280.2021.9342280} emphasized the heartbeat’s higher harmonics by using a differential enhancement method that combines the first derivative (i.e., $(\dd/\dd t) \angle s(t)$) with the second derivative (i.e., $(\dd^2/\dd t^2) \angle s(t)$) of the radar signal’s phase. 
Nosrati and Tavassolian \cite{https://doi.org/10.1109/JERM.2018.2879452} used the second derivative of a complex radar signal (i.e., $(\dd^2/\dd t^2) s(t)$) to emphasize the heartbeat component in a method where the complex signal is used instead of its phase.
Sakamoto and Yamaguchi \cite{Yamaguchi2019} used the absolute value of the second derivative of the complex radar signal (i.e., they used $|(\dd^2/\dd t^2) s(t)|$) to estimate the accuracy when measuring the heart rate.

In this study, we demonstrate the effectiveness of using the absolute value of the second derivative of the complex radar signal, which was proposed previously in \cite{https://doi.org/10.1109/JERM.2018.2879452,Yamaguchi2019}, when emphasizing the heartbeat’s higher harmonics and estimating the heart rate. The advantage of this approach is that the method is not dependent on the phase demodulation accuracy because the complex signal is used directly in the estimation process. This study also introduces the variational mode extraction (VME) process, which is a mode decomposition method, to enable further amplification of the heartbeat component that is emphasized by the differentiation process. The performance of the proposed method is evaluated quantitatively by conducting radar measurement experiments that involve eleven participants.

\section{Radar Measurement of Physiological Signals}\label{section.2}
\subsection{Radar Imaging and Estimation of Body Displacement}
In this section, we use a 79 GHz frequency-modulated continuous-wave (FMCW) radar system with array antennas.
Using the FMCW demodulation method and array signal processing, the complex radar image $I'_\mathrm{C}(r,\theta,t)$ is obtained, where $r$, $\theta$, and $t$ are the range, the azimuth angle, and the slow time, respectively. To remove any stationary clutter, the time-average is subtracted from $I'_\mathrm{C}(r,\theta,t)$ as $I_\mathrm{C}(r, \theta,t) = I'_\mathrm{C}(r, \theta,t)-({1}/{T})\int_0^TI'_\mathrm{C}(r, \theta,t)\dd t$ to generate $I_\mathrm{C}(r,\theta,t)$, where $T$ is the measurement time. We then generate a power radar image $I_\mathrm{P} = |I_\mathrm{C}|^2$, which is time-averaged to give the time-averaged power radar image $I_\mathrm{A}(r, \theta)=({1}/{T})\int_{0}^{T}I_\mathrm{P}(r, \theta,t)\dd t$. The target position $(r_0,\theta_0)$ is then determined for estimation of the physiological signals using $I_{\mathrm{A}}(r,\theta)$ and we obtain the complex radar signal $s(t)=I_\mathrm{C}(r_0,\theta_0,t)$ along with the body displacement $d(t)$, where $d(t) = (\lambda/4\pi)\angle s(t)$ and $\lambda$ and $\angle$ denote the wavelength and phase of a complex number, respectively.

\subsection{Mode Decomposition Methods}
Mode decomposition, which is a type of blind source separation method, can decompose an input signal into multiple components and then reconstruct the signal by simply selecting the desired mode~\cite{https://doi.org/10.1109/TIM.2023.3300471, https://doi.org/10.1145/3627161}. Therefore, it is expected that application of mode decomposition to the proposed $|s''(t)|$ will cause the heartbeat harmonic components to be enhanced, thus leading to an improvement in the accuracy when estimating IBIs. Mode decomposition has been applied previously to the phase $\psi(t)$ or the displacement $d(t)$ for estimation of the respiration and heartbeat parameters \cite{https://doi.org/10.1049/joe.2019.0619, https://doi.org/10.1145/3556558.3558578, https://doi.org/10.1109/TBME.2013.2288319, https://doi.org/10.1109/ACCESS.2020.2985286, https://doi.org/10.1109/JSEN.2021.3075109}.

VMD~\cite{https://doi.org/10.1109/TSP.2013.2288675}, which is a type of blind source separation method, was developed by Dragomiretskiy and Zosso in 2014. In VMD, the intrinsic mode function (IMF) is defined as an amplitude- and frequency-modulated signal:
\begin{align}
    u_k(t) = A_k(t)\cos(\phi_k(t)),
\end{align}
where $k = 1, \ldots,K_\mathrm{v}$ and $K_\mathrm{v}$ is the number of modes. The phase $\phi_k(t)$ is a nondecreasing function, i.e., $\phi'_k(t) = (\dd/\dd t)\phi_k(t)\geq 0$. The envelope $A_k(t)\geq 0$ and both $A_k(t)$ and the instantaneous frequency $\phi'_k(t)$ vary sufficiently slowly such that, over a sufficiently long time $[t-t_0, t+t_0]$ (where $t_0 \geq  2\pi/\phi'_k(t)$), $u_k(t)$ can be considered to be a pure harmonic signal with amplitude $A_k(t)$ and an instantaneous frequency $\phi'_k(t)$. Therefore, the IMF has a specific sparsity property~\cite{https://doi.org/10.1109/TSP.2013.2288675}. In VMD, each IMF $u_k$ is assumed to located around the $k$th central frequency $f_k$. Finding the pairing of $u_k$ and $f_k$ that minimizes the constrained variational problem allows the real-valued input signal to be decomposed into each mode.

The VMD is superior to empirical mode decomposition (EMD) methods in terms of both IMF separation and robustness against noise \cite{https://doi.org/10.1016/j.ymssp.2015.02.020}. Because of its high performance, VMD has been used widely in radar measurements of physiological signals \cite{https://doi.org/10.1109/TIM.2023.3267348, https://doi.org/10.1109/JIOT.2021.3075167, https://doi.org/10.1109/JSEN.2022.3148003, https://doi.org/10.1109/EMBC46164.2021.9630414, https://doi.org/10.3390/ani10020205, https://doi.org/10.1007/s11517-022-02678-x}.
Lele et al. reconstructed both respiratory and heartbeat signals by summing IMFs that satisfied a threshold ratio condition between the total energy of each IMF in the frequency domain and the energy contained within the fundamental frequency ranges of the respiration and heartbeat signals only \cite{https://doi.org/10.1109/Radar53847.2021.10028620}.
However, the VMD performance is dependent on the number of decomposition layers $K_\mathrm{v}$ used and the penalty coefficient $\alpha$~\cite{https://doi.org/10.1109/TSP.2013.2288675, https://doi.org/10.1109/CIEEC54735.2022.9845851}. A low penalty coefficient $\alpha$ can lead to reduced decomposition accuracy because of noise. The number of decomposition layers $K_\mathrm{v}$ used also affects the decomposition results; a small $K_\mathrm{v}$, which means fewer IMFs, can cause mode mixing. A large $K_\mathrm{v}$ can lead to overlapping of the central frequencies, decomposition of the signals into more than two IMFs, and reduced mode decomposition accuracy.

In this study, we use VME~\cite{https://doi.org/10.1109/JBHI.2017.2734074}, a method based on VMD that was proposed by Nazari and Sakhaei, to extract the heartbeat harmonic components effectively. The VME was developed from VMD to extract the respiratory signals from electrocardiograph (ECG) signals~\cite{https://doi.org/10.1109/JBHI.2017.2734074}. In VME, the desired mode $ u_\mathrm{d}(t) $ and the residual $ r_\mathrm{d}(t) $ are defined, and a penalty term is then added to the VMD objective function to define the optimization problem as follows: 
\begin{align}
     & \underset{u_\mathrm{d}, \, \omega_\mathrm{d}, \,r_\mathrm{d}}{\min}\left\lbrace \alpha B_\mathrm{d}^2 + \Vert\beta(t)\ast r_\mathrm{d}(t)\Vert_2^2\right\rbrace \\
     & \mathrm{\quad s.\,t.}\quad u_\mathrm{d}(t) + r_\mathrm{d}(t) = |s''(t)|,
    \label{eq:VME_mokuteki}
\end{align}
where $\alpha$ is a parameter that balances the first and second terms of the objective function, $B_\mathrm{d}$ is the bandwidth of the desired mode, and $\beta(t)$ is the impulse response of a filter, which is described by 
\begin{align}
    B(f) = \mathcal{F}\lbrace\beta(t)\rbrace = \frac{1}{4\pi^2 \alpha(f-f_\mathrm{d})^2}
\end{align}
and has infinite gain at $ f = f_\mathrm{d} $, but it behaves like a Wiener filter at frequencies other than $f_\mathrm{d}$.
Unlike VMD, which determines all modes from the input signal and then decomposes them iteratively into each mode, VME extracts only a specific mode $u_\mathrm{d}(t)$, which has a known approximate central frequency $f_\mathrm{d}$. Therefore, when compared with VMD, VME has both a higher convergence rate and a lower computational load. In addition to these characteristics, VME is a useful method for applications in which specific modes are extracted because it does not need to provide $ K_\mathrm{v} $~\cite{https://doi.org/10.1109/ACCESS.2023.3302918, https://doi.org/10.1109/JBHI.2017.2734074}.
Similar to VMD, VME is affected by the parameter $ \alpha $; if the value of $\alpha$ is too high, then too many modes with narrow bandwidths will be obtained, leading to generation of artifacts due to noise; if the value of $\alpha$ is too low, then there is the possibility that multiple modes will be mixed, thus leading to the issue where the desired signal is not recognized as a single mode. Although a method for optimization of the VME parameters using a genetic algorithm has been proposed by Zhang et al.~\cite{https://doi.org/10.1109/SCSET58950.2023.00172}, $ \alpha $ is determined empirically in this study.

\section{Proposed Heart Rate Estimation Method}\label{section.3}
Although many existing studies have used the signal phase $\psi(t)$ or the displacement $d(t)$ to measure the heartbeat \cite{https://doi.org/10.1109/TIM.2023.3300471, https://doi.org/10.1109/JPROC.2023.3244362, https://doi.org/10.1109/JERM.2022.3214753}, the accuracy obtained is not necessarily satisfactory because of the interference that occurs with the respiratory components. In this study, we propose a method that uses the absolute value of the second derivative of the complex radar signal $|(\dd^2/\dd t^2)s(t)|=|s''(t)|$ rather than the displacement $\psi(t)$, as described in \cite{https://doi.org/10.1109/JERM.2018.2879452,Yamaguchi2019}.

\begin{figure}[tb]
    \centering
    \begin{minipage}{0.45\linewidth}
        \centering
        \includegraphics[width=\linewidth]{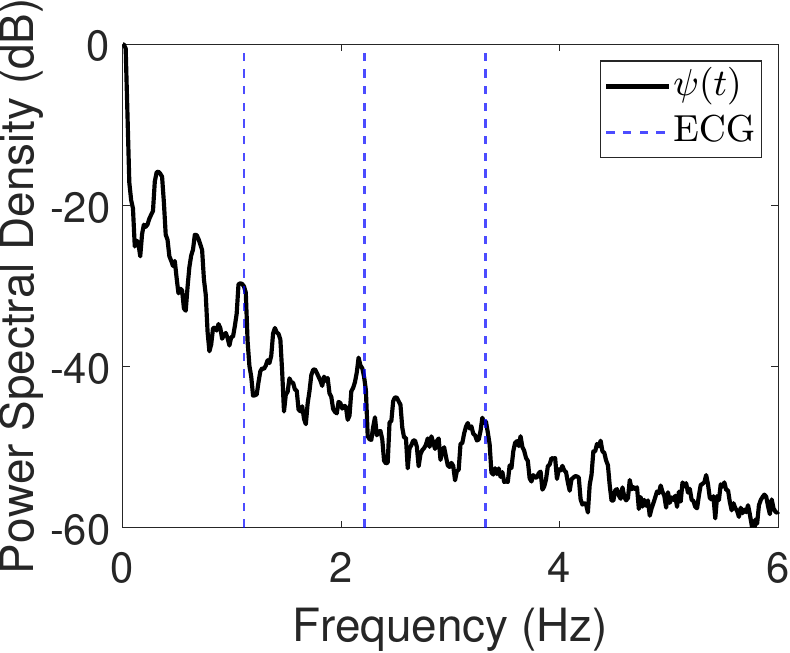} 
        \subcaption{}
        \label{fig:subA_disp_fft}
    \end{minipage}
    \begin{minipage}{0.45\linewidth}
        \centering
        \includegraphics[width=\linewidth]{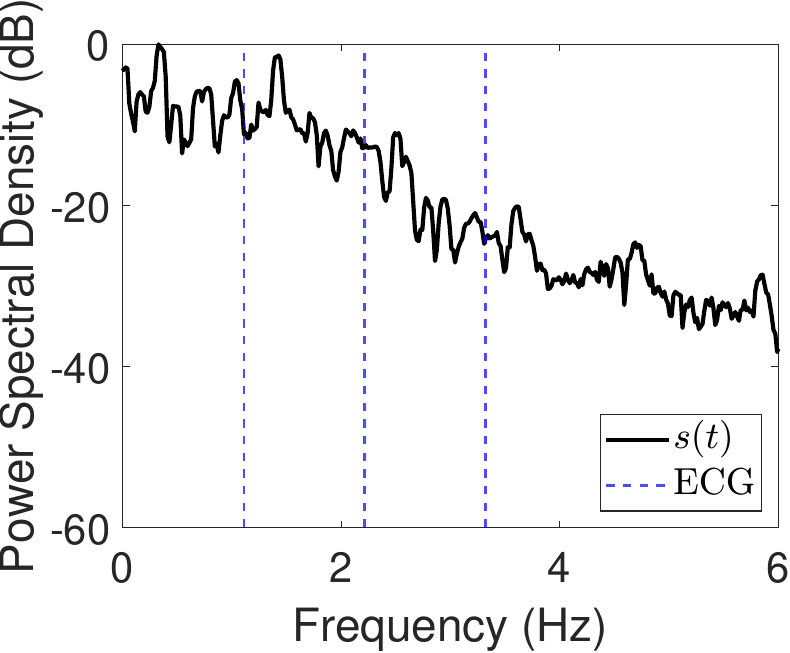} 
        \subcaption{}
        \label{fig:subA_Ic_fft}
    \end{minipage}
    \begin{minipage}{0.45\linewidth}
        \centering
        \includegraphics[width=\linewidth]{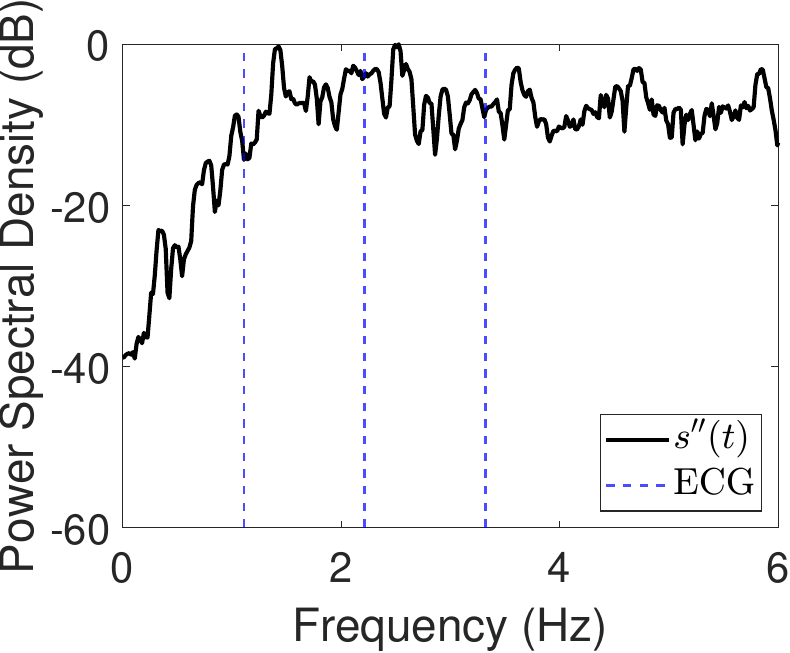} 
        \subcaption{}
        \label{fig:subA_diffIc_fft}
    \end{minipage}
    \begin{minipage}{0.45\linewidth}
        \centering
        \includegraphics[width=\linewidth]{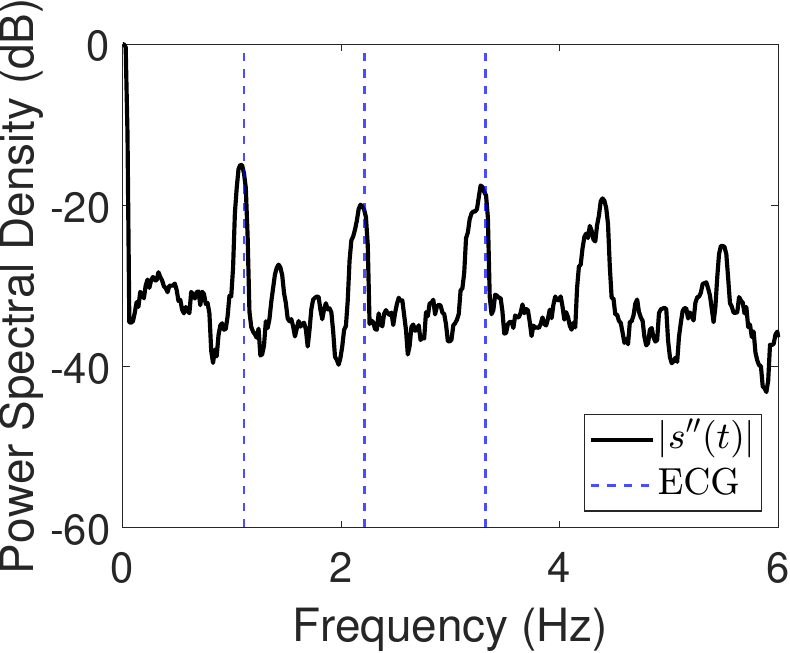} 
        \subcaption{}
        \label{fig:subA_sacc_fft}
    \end{minipage}
    \caption{Participant A's power spectral density functions for various forms of radar signal (black lines), along with the fundamental and second and third harmonic frequencies of the heartbeat signal obtained from an ECG. (a) $|\mathcal{F}\lbrace \psi(t)\rbrace|^2$, (b) $|\mathcal{F}\lbrace s(t)\rbrace|^2$, (c) $|{\mathcal F}\lbrace s''(t)\rbrace|^2$, and (d) the proposed |$\mathcal{F}\lbrace |s''(t)|\rbrace|^2$.}
    \label{fig:PeriodogramA}
\end{figure}

\begin{figure}[tb]
    \centering
    \begin{minipage}{0.45\linewidth}
        \centering
        \includegraphics[width=\linewidth]{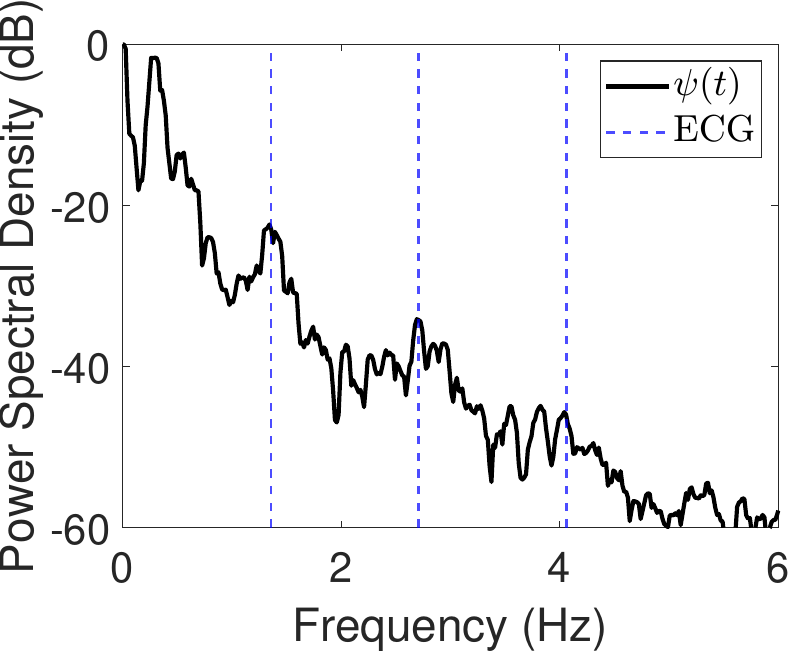} 
        \subcaption{}
        \label{fig:subB_disp_fft}
    \end{minipage}
    \begin{minipage}{0.45\linewidth}
        \centering
        \includegraphics[width=\linewidth]{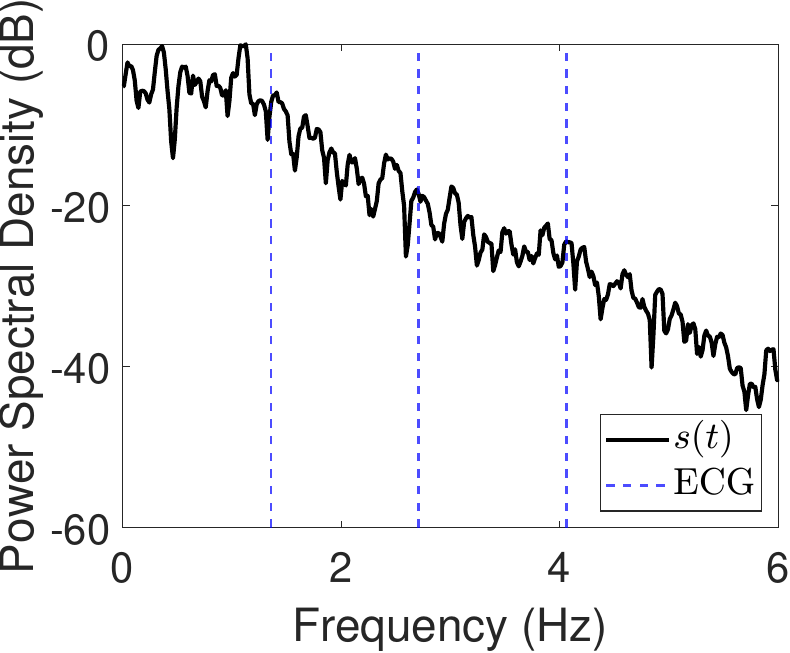} 
        \subcaption{}
        \label{fig:subB_Ic_fft}
    \end{minipage}

    \begin{minipage}{0.45\linewidth}
        \centering
        \includegraphics[width=\linewidth]{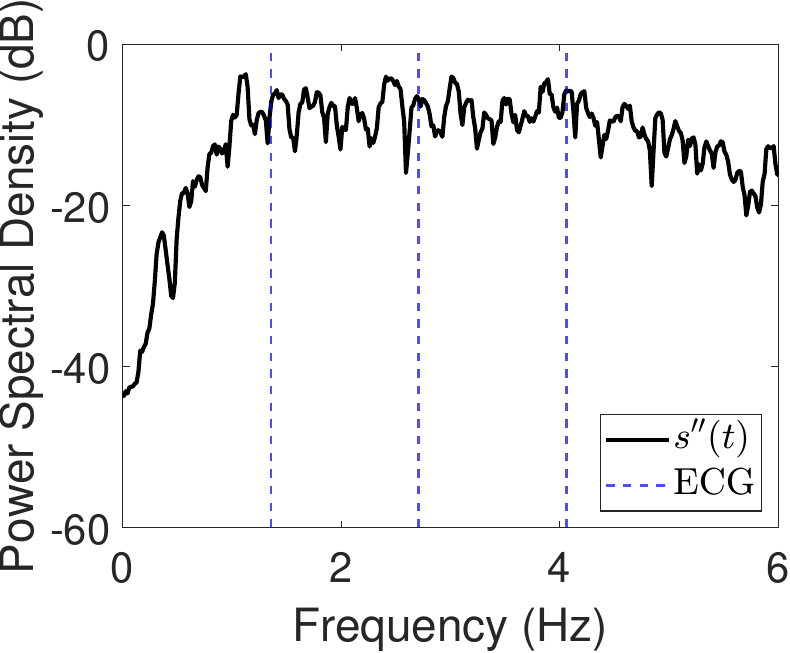} 
        \subcaption{}
        \label{fig:subB_diffIc_fft}
    \end{minipage}
    \begin{minipage}{0.45\linewidth}
        \centering
        \includegraphics[width=\linewidth]{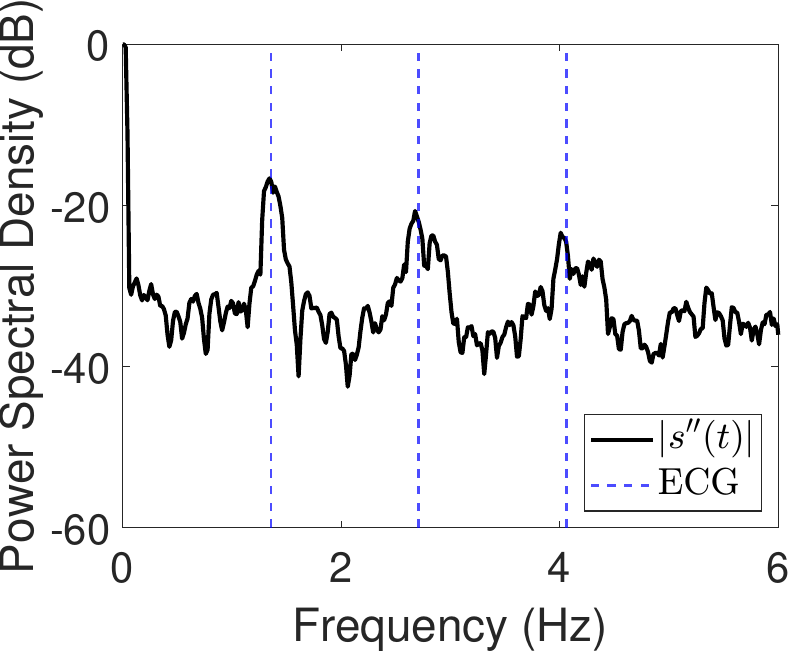} 
        \subcaption{}
        \label{fig:subB_sacc_fft}
    \end{minipage}

   \caption{Participant B's power spectral density functions for various forms of radar signal (black lines), along with the fundamental and second and third harmonic frequencies of the heartbeat signal obtained from an ECG. (a) $|\mathcal{F}\lbrace \psi(t)\rbrace|^2$, (b) $|\mathcal{F}\lbrace s(t)\rbrace|^2$, (c) $|{\mathcal F}\lbrace s''(t)\rbrace|^2$, and (d) the proposed $|\mathcal{F}\lbrace |s''(t)|\rbrace|^2$.}
   \end{figure}

Here, we present examples of signals that have been processed using different methods to demonstrate the effectiveness of our proposed approach, with radar signals that were measured from two participants (A and B) being used for this purpose. Details of the experiments are provided in Section IV. First, we consider the power spectrum of $\psi(t)= \mathrm{unwrap}\{\angle s(t)\}$, where $\mathrm{unwrap}$ is a phase unwrapping function. The power spectra of $\psi(t)$ for participants A and B are shown in Fig. \ref{fig:subA_disp_fft} and Fig. \ref{fig:subB_disp_fft}, respectively. Please note that the displacement $d(t)$ can be estimated as $d(t)=(\lambda/4\pi)\psi(t)$ if the signal contains only a single echo that has been reflected from a target with a time-varying displacement. The fundamental frequency and the second and third harmonics of the heartbeat that were obtained from the ECG are shown as blue dashed lines. In these figures, we see peaks that correspond to the fundamental and harmonic components of the heartbeat in addition to the much larger peaks that correspond to the respiratory components. The accuracy of heart rate estimation from $d(t)$ is dependent on the intensity and frequency of the heartbeat and respiration signals; as a result, the accuracy is degraded in some cases.

Next, we consider the power spectrum of the complex radar signal $s(t)$ itself; the corresponding spectra are shown in Fig.~\ref{fig:subA_Ic_fft} and Fig.~\ref{fig:subB_Ic_fft} for participants A and B, respectively. In these figures, the peaks that correspond to the heartbeat components are not shown clearly, which indicates that it is difficult to estimate the heart rate from $|\mathcal{F}\lbrace s(t)\rbrace|^2$ at least partly because of the interference from the respiratory harmonic components and the nonlinear effect of the phase modulation. To emphasize the heartbeat harmonics at higher frequencies, we consider the power spectrum of the second derivative of the complex radar signal $s''(t)$; the corresponding spectra are shown in Fig.~\ref{fig:subA_diffIc_fft} and Fig.~\ref{fig:subB_diffIc_fft} for participants A and B, respectively. Comparison of Fig.~\ref{fig:subA_Ic_fft} with Fig.~\ref{fig:subA_diffIc_fft} and Fig.~\ref{fig:subB_Ic_fft} with Fig.~\ref{fig:subB_diffIc_fft} shows that the high-frequency components have been emphasized by the derivative operation. Despite this emphasis, we are still unable to see clear peaks corresponding to the heartbeat components, even after differentiation. 

Fig. \ref{fig:subA_sacc_fft} and Fig. \ref{fig:subB_sacc_fft} show the power spectral density characteristics that correspond to $\mathcal{F}\lbrace |s''(t)|\rbrace$ for participants A and B, respectively. When compared with Figs. \ref{fig:subA_Ic_fft} and \ref{fig:subB_Ic_fft}, we see that Figs. \ref{fig:subA_sacc_fft} and \ref{fig:subB_sacc_fft} both show clear peaks that correspond to the heartbeat’s fundamental and harmonic components. Additionally, comparison of Fig. \ref{fig:subA_sacc_fft} and Fig. \ref{fig:subB_sacc_fft} with Fig. \ref{fig:subA_disp_fft} and Fig. \ref{fig:subB_disp_fft} shows that the heartbeat components, including the harmonic components, are emphasized in the former cases. These examples demonstrate the effectiveness of using the absolute value of the second derivative of the complex radar signal $|s''(t)|$ when measuring heartbeat signals using radar. 

\begin{figure}[tb]
    \centering
    \begin{minipage}{0.45\linewidth}
        \centering
        \includegraphics[width=\linewidth]{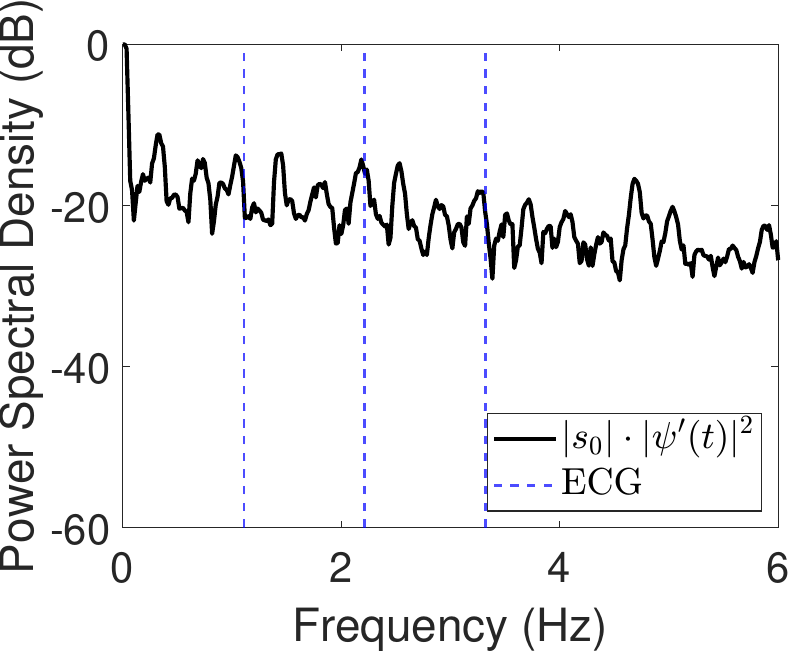} 
        \subcaption{}
        \label{fig:A_diff1}
    \end{minipage}
    \begin{minipage}{0.45\linewidth}
        \centering
        \includegraphics[width=\linewidth]{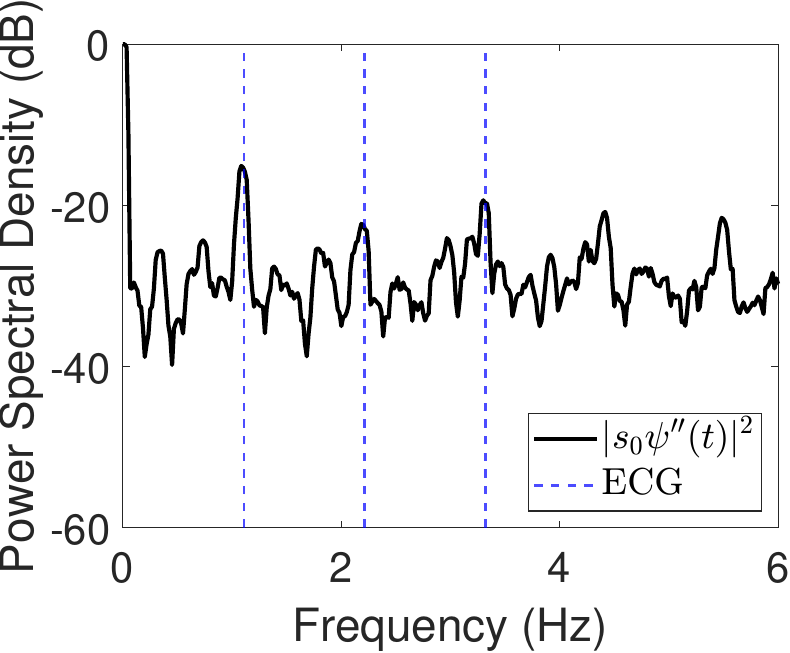} 
        \subcaption{}
        \label{fig:A_diff2}
    \end{minipage}

    \caption{Participant A's power spectra for various forms of radar signal (black lines), along with the fundamental and second and third harmonic frequencies of the heartbeat signal obtained from an ECG; (a) $|{s''}_1(t)| = |s_0|\cdot |\psi'(t)|^2$ and (b) $|{s''}_2(t)| = |s_0 \psi''(t)|$.}
    \label{fig:SubADerivativeFirstAndSecond}
\end{figure}

\begin{figure}[tb]
    \centering
    \begin{minipage}{0.45\linewidth}
        \centering
        \includegraphics[width=\linewidth]{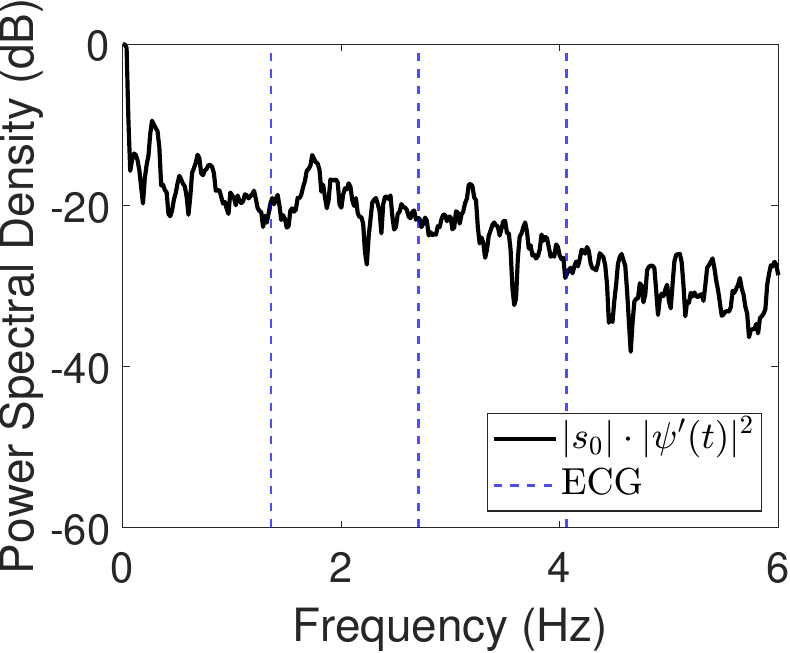} 
        \subcaption{}
        \label{fig:B_diff1}
    \end{minipage}
    \begin{minipage}{0.45\linewidth}
        \centering
        \includegraphics[width=\linewidth]{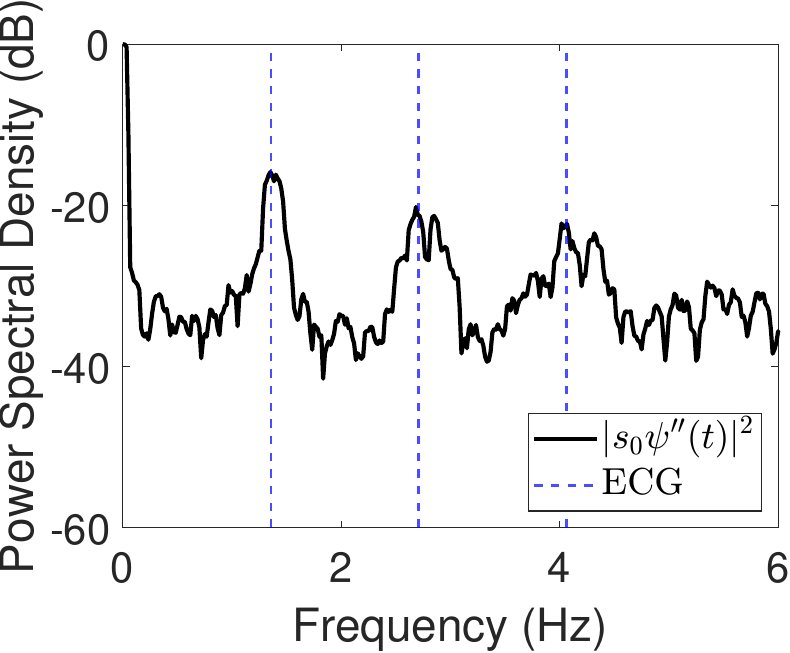} 
        \subcaption{}
        \label{fig:B_diff2}
    \end{minipage}

    \caption{Participant B's power spectra for various forms of radar signal (black lines), along with the fundamental and second and third harmonic frequencies of the heartbeat signal obtained from an ECG; (a) $|{s''}_1(t)| = |s_0|\cdot |\psi'(t)|^2$ and (b) $|{s''}_2(t)| = |s_0 \psi''(t)|$.}
    \label{fig:SubBDerivativeFirstAndSecond}
\end{figure}

Here, we discuss which of the terms contributes to this emphasis of the heartbeat components. For the complex radar signal $s(t)$ that is expressed as $s(t)=s_0(t)\ee^{\jj \psi(t)}$ and has the amplitude $s_0(t)$, which is an almost constant complex amplitude (i.e., $(\dd/\dd t)s_0(t)\simeq 0$), and a phase $\psi(t)$ that is a real function of $t$, the absolute value of the second derivative of $|s''(t)|$ can be expressed as
\begin{align}
    \left\vert\frac{\dd^2 }{\dd t^2}s(t)\right\vert                                                                                                           & \simeq |s_0(t)|\left\vert\left\lbrace - \left(\frac{\dd }{\dd t}\psi(t)\right)^2 +\mathrm{j}\frac{\dd^2 }{\dd t^2}\psi(t) \right\rbrace \ee^{\mathrm{j}\psi(t)}\right\vert \\
                      & = |s_0(t)| \sqrt{\left|\psi'(t)\right|^4 + \left|\psi''(t)\right|^2}, \label{eq:signal_acc}
\end{align}
which comprises $\psi'(t)$ and $\psi''(t)$, which are the first and second derivatives of the phase $\psi(t)$, respectively. 
Let us evaluate and compare the effectiveness of $\psi'(t)$ and $\psi''(t)$ as follows; by substituting $\psi''(t)=0$ into Eq. (\ref{eq:signal_acc}), we obtain $|{s''}_1(t)| = |s_0(t)|\cdot |\psi'(t)|^2$; and by substituting $\psi'(t)=0$ into Eq. (\ref{eq:signal_acc}), we obtain $|{s''}_2(t)| = |s_0(t) \psi''(t)|$. The power spectra of $|{s''}_1(t)|$ and $|{s''}_2(t)|$ are shown in Fig. \ref{fig:A_diff1} and \ref{fig:A_diff2} for participant A and are shown in Fig. \ref{fig:B_diff1} and \ref{fig:B_diff2} for participant B, respectively. Please note that, in these power spectra, $s_0(t)$ is a function of time and is dependent on $t$, unlike the case in Eq. (\ref{eq:signal_acc}). From these figures, we can see that the heartbeat component is emphasized in the case of $|{s''}_2(t)|$ in particular because it contains the second derivative of the phase $\psi''(t)$.

Furthermore, when the power spectra of $|{s''}_2(t)|$ shown in Figs. \ref{fig:A_diff2} and \ref{fig:B_diff2} are compared with the results from the proposed $|s''(t)|$ in Figs. \ref{fig:subA_sacc_fft} and \ref{fig:subB_sacc_fft}, we can see that the proposed approach is more effective than use of $|{s''}_2(t)|$, thus indicating the superior performance of the proposed approach. Another advantage of the proposed approach based on $|s''(t)|$ is that it allows the phase unwrapping process to be skipped. Given that the phase unwrapping does not work correctly when the signal is noisy, this advantage is crucial when the method is applied in practice.

\begin{figure}[tb]
    \centering
    \includegraphics[width = 0.7\linewidth,pagebox=cropbox,clip]{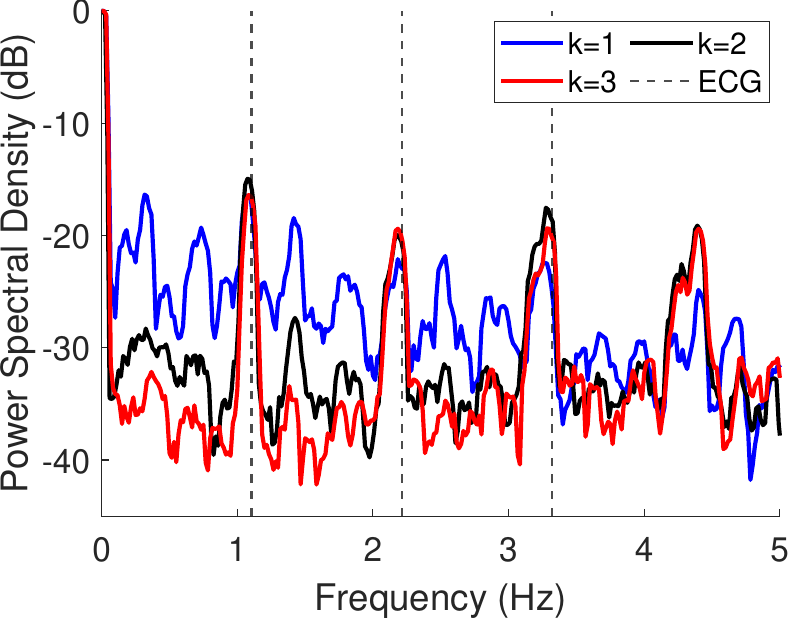}
    \caption{Participant A's power spectra $|\mathcal{F}\lbrace\vert s^{(k)}(t)\vert\rbrace|^2$, where the dashed lines indicate the fundamental and harmonic frequencies of the heartbeat.}
    \label{fig:subA_addacc_fft}
\end{figure}

\begin{figure}[tb]
    \centering
    \includegraphics[width = 0.7\linewidth,pagebox=cropbox,clip]{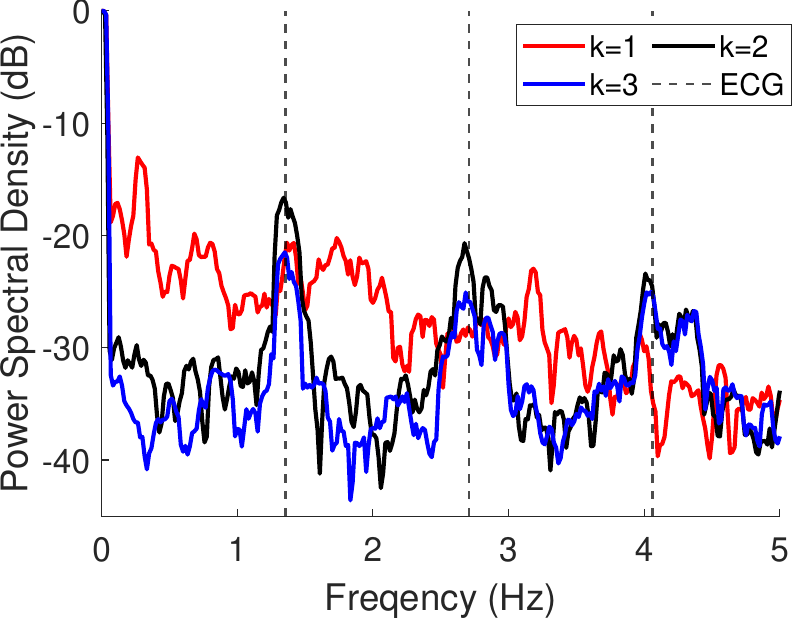}
    \caption{Participant B's power spectra $|\mathcal{F}\lbrace\vert s^{(k)}(t)\vert\rbrace|^2$, where the dashed lines indicate the fundamental and harmonic frequencies of the heartbeat.} 
    \label{fig:subB_addacc_fft}
\end{figure}

The power spectra given by 
\begin{equation}
|\mathcal{F}\lbrace\vert s^{(k)}(t)\vert\rbrace|^2 = \left|\mathcal{F}\left\lbrace \left|\frac{\dd^k}{\dd t^k} s(t) \right|\right\rbrace
\right|^2\;\; (k = 1, \ldots,3)
\end{equation}
for participant A are shown in Fig. \ref{fig:subA_addacc_fft} and those for participant B are shown in Fig. \ref{fig:subB_addacc_fft}. Figs. \ref{fig:subA_addacc_fft} and \ref{fig:subB_addacc_fft} show that the heartbeat component is emphasized most strongly for $k = 2$, which justifies use of our proposed approach with $|s''(t)|$. 

Please note that Nosrati and Tavassolian \cite{https://doi.org/10.1109/JERM.2018.2879452} also examined the characteristics for $|\mathcal{F}\lbrace s^{(k)}(t)\rbrace|^2$ without taking the absolute value, instead of using our proposed approach with $|\mathcal{F}\lbrace |s^{(k)}(t)|\rbrace|^2$, and also demonstrated the effectiveness of the differentiation for $k=2$ among the other values of $k$. They also noted that when the order of the differentiation increased, the harmonics of the respiration and the intermodulation waves of the respiration and heartbeat were also amplified, which reduced the signal-to-noise ratio (SNR) of the heartbeat component in addition to the numerical instability that was caused by the differentiation process. As they suggested, the heartbeat’s SNR is expected to decrease when $k\geq 3$ because the noise in the higher frequency band is also emphasized by the differentiation operation. Therefore, we proposed use of the absolute value of the second derivative $|s''(t)|$. 


The definitions for the conventional and proposed methods are summarized in Table \ref{tbl:Sakamoto}. The conventional methods all use $\psi(t)$, which is the unwrapped phase of the complex radar signal, whereas the proposed methods all use $|s''(t)|$, which is the absolute value of the second derivative of the complex radar signal. Conventional method 1 and proposed method 1 do not apply VME; conventional method 2 and proposed method 2 both apply VME to extract the desired mode $u_\mathrm{d}(t)$ by using the desired frequency $f_\mathrm{d}$, which corresponds to the second harmonic frequency of the heartbeat signal. Here, $f_\mathrm{d}$ is selected as the frequency that corresponds to the maximum power spectrum density within the frequency range from 2.0--3.4 Hz; i.e., $f_\mathrm{d}=\arg\max_{f} |\mathcal{F}\lbrace |s''(t)|\rbrace|^2$. Please note that if a local maximum point does not exist within the frequency range, we then set $f_\mathrm{d}=2.7$ Hz, which is the center frequency of the frequency range. Conventional method 3 and proposed method 3 follow the same steps used in conventional method 2 and proposed method 2, respectively, to extract $u_\mathrm{d,1}(t)$; then they apply the VME again using the desired frequency $(3/2)f_\mathrm{d}$, which corresponds to the third harmonic frequency, extract the desired mode $u_\mathrm{d,2}(t)$, and subsequently use the summation $u_\mathrm{d,1}(t)+u_\mathrm{d,2}(t)$. Conventional methods 1, 2, and 3, and proposed methods 1, 2, and 3 then all apply the topology method \cite{Sakamoto2016} to estimate the IBI. The VME parameter $\alpha$ is set empirically at values of $\alpha=10^5$ and $\alpha=3\times 10^4$ for the conventional and proposed methods, respectively.

\begin{table}[tb]
   \centering
    \caption{CONVENTIONAL AND PROPOSED METHODS FOR IBI ESTIMATION}
\begin{tabular}{|c|l|c|c|}
\hline
Method  & \multicolumn{1}{c|}{Signal to Process}                                                                 & \begin{tabular}[c]{@{}c@{}}Mode \\ Decomposition\end{tabular} & \begin{tabular}[c]{@{}c@{}}IBI\\ Estimation\end{tabular}    \\ \hline
Conv. 1 & \multirow{3}{*}{\begin{tabular}[c]{@{}l@{}}$\psi(t)$\\ $=\mathrm{unwrap}\{\angle s(t)\}$\end{tabular}} & ---                                                           & \multirow{6}{*}{\begin{tabular}[c]{@{}c@{}}Topology \\ method\end{tabular}} \\ \cline{1-1} \cline{3-3}
Conv. 2 &                                                                                                        & VME mode 1                                                    &                                                                             \\ \cline{1-1} \cline{3-3}
Conv. 3 &                                                                                                        & VME modes 1 \& 2                                               &                                                                             \\ \cline{1-3}
Prop. 1 & \multirow{3}{*}{$\displaystyle \left|\frac{\dd^2}{\dd t^2}s(t)\right|$}                                              & ---                                                           &                                                                             \\ \cline{1-1} \cline{3-3}
Prop. 2 &                                                                                                        & VME mode 1                                                    &                                                                             \\ \cline{1-1} \cline{3-3}
Prop. 3 &                                                                                                        & VME modes 1 \& 2                                               &                                                                             \\ \hline
\end{tabular}
    \label{tbl:Sakamoto}
\end{table}

\section{Performance Evaluation of the Proposed Methods}\label{section.4}
\begin{figure}[tb]
    \centering
    \includegraphics[width = 0.7\linewidth]{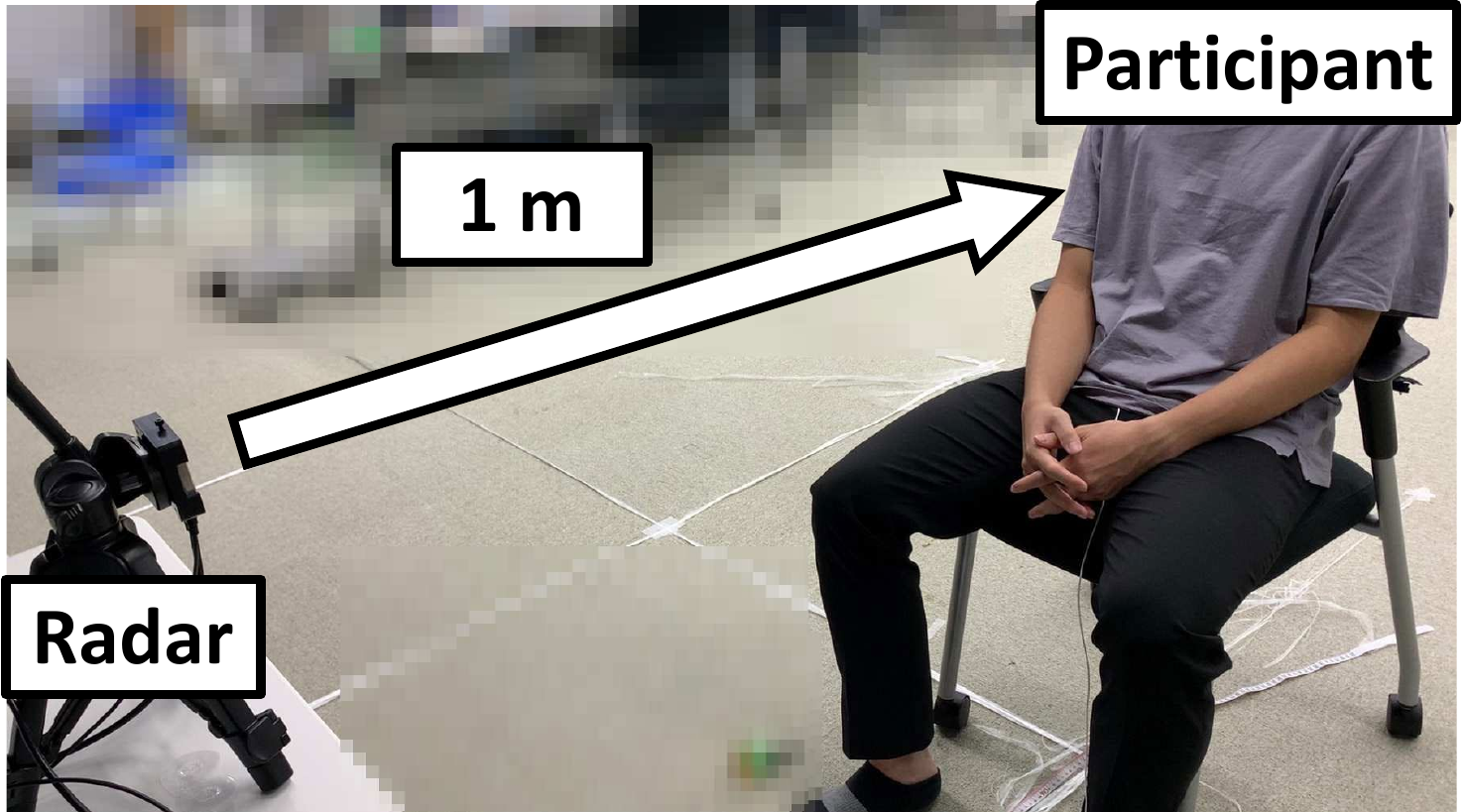}
    \caption{Experimental scene for the measurement of a participant while breathing normally using millimeter wave radar.}
    \label{fig:acc_vmd_experi_photo}
\end{figure}

We used a 79 GHz millimeter-wave FMCW radar system with an antenna array composed of three transmitting elements and four receiving elements; the array specifications are listed in Table \ref{tbl:radar_system}. The element intervals were $ 2\lambda $ and $ \lambda/2 $ for the transmitting and receiving elements, respectively. The radar system was located at a distance of $1.0$ m away from each participant.
The radar measurements were performed for eleven participants who were all healthy adults, and a total of 25 samples of these measurements were used to evaluate the corresponding performances of the conventional and proposed methods. All participants were seated while breathing naturally for 60.0 s and the experimental scene is shown in Fig. \ref{fig:acc_vmd_experi_photo}. An ECG sensor was attached to each participant to evaluate the accuracy of estimation of the heart IBIs.

\begin{table}[tb]
    \centering
    \caption{SPECIFICATIONS OF RADAR SYSTEM}
    \begin{threeparttable}[h]
    \begin{tabular}{l|cc}
        \toprule
                                            & Specification                               & unit                      \\
        \midrule
        Tx element spacing    & 7.6                  & mm                        \\
        Output power (EIRP)\tnote{*}                 & 23                                & dBm                       \\
        Tx element beamwidth          & \multirow{2}{*}{$\pm$35 / $\pm$4} & \multirow{2}{*}{deg /deg} \\(azimuth/elevation)            &&     \\
        Rx element spacing   & 1.9                 & mm                        \\
        Rx element beamwidth & \multirow{2}{*}{$\pm$35 / $\pm$4} & \multirow{2}{*}{deg /deg} \\(azimuth/elevation)               & &\\
        Frequency range                  & 77.0 -- 80.9                      & GHz                       \\
        Sampling intervals      & 6.87                              & ms                        \\
        \bottomrule
    \end{tabular}
    \begin{tablenotes}
        \item[*]EIRP: Equivalent Isotropically Radiated Power
    \end{tablenotes}
    \end{threeparttable}
        \label{tbl:radar_system}
\end{table}

The performance of each method is evaluated in terms of the correlation coefficient (CC) between the actual IBI ($h_0(t)$) and the estimated IBI ($h(t)$), as given by 
\begin{equation}
\frac{\int_0^T h_0(t)h(t)\dd t}
{\sqrt{\int_0^T h_0^2(t)\dd t}\sqrt{\int_0^T h^2(t)\dd t}},
\end{equation}
and the root-mean-square error (RMSE), as given by 
\begin{equation}
\sqrt{\frac{1}{T}\int_0^T |h(t)-h_0(t)|^2\dd t}.
\end{equation}
The performance is also evaluated in terms of the time coverage rate (TCR) \cite{https://doi.org/10.1587/transcom.2020EBP3078}; the TCR is defined here as $n/N$, where $N$ is the total number of time segments, which means that $N=T/T_0$, where $T_0$ is the time segment length; in addition, $n$ is the number of time segments in which at least one IBI exists with an absolute error of less than $T_\mathrm{th}$ (i.e., $\exists t \in T_n \;\;\;\; \vert h(t) - h_0(t) \vert \leq T_\mathrm{th}$ for $T_n = [nT_0, (n+1)T_0]$. We set these parameters empirically at $T_0 = 0.5 $ s and $ T_\mathrm{th} = 50 $ ms. Please note that an accurate method is expected to achieve a high CC, a high TCR, and a low RMSE.

\begin{table}[tb]
    \centering
    \caption{IBI ESTIMATION EVALUATION INDEX AVERAGES FOR EACH METHOD}
    \resizebox{\linewidth}{!}{\begin{tabular}{c|rr|rr|rr}
        \toprule
        \multirow{2}{*}{} & \multicolumn{6}{c}{Method}                                                                                                                                                                                                               \\
                                  & \multicolumn{1}{c}{Conv. 1} & \multicolumn{1}{c}{Prop. 1} & \multicolumn{1}{c}{Conv. 2} & \multicolumn{1}{c}{Prop. 2} & \multicolumn{1}{c}{Conv. 3} & \multicolumn{1}{c}{Prop. 3} \\\midrule
        CC              & {0.51}                   & {0.64}                   &{0.47}              & {0.66}              & {0.47}                 & {0.74}               \\
        RMSE (ms)                 & 47.07                                   & 33.42                                   & 37.03                              & 31.44                              & 34.42                                 & 26.02                               \\
        TCR (\%)                  & 37.10                                   & 38.19                                   & 61.83                              & 77.61                              & 69.01                                 & 88.27                               \\
        \bottomrule
    \end{tabular}
    }
    \label{tbl:acc_vmd_result_ave}
\end{table}

Table \ref{tbl:acc_vmd_result_ave} shows the average CC, RMSE, and TCR values for each method. Comparison of proposed methods 1, 2, and 3 shows that proposed method 3 achieves the best improvement in terms of the CC, TCR and RMSE values, thus indicating the effectiveness of the approach based on extraction of the second and third heartbeat harmonics using VME. These results indicate that the topology method benefits from the heartbeat harmonics of orders higher than three, rather than from signals that contain only the second heartbeat harmonic component, and also confirm that use of $|s''(t)|$ allows higher accuracy to be achieved when compared with the conventional methods that use $\psi(t)$, which is proportional to the displacement $d(t)$. When the proposed method was used, the CC increased by 0.20, the RMSE decreased by 23\%, and the TCR increased by 19\% on average, thus demonstrating the effectiveness of the proposed method.

\begin{figure}[tb]
    \centering
    \begin{minipage}{0.49\linewidth}
        \centering
        \includegraphics[width = \linewidth]{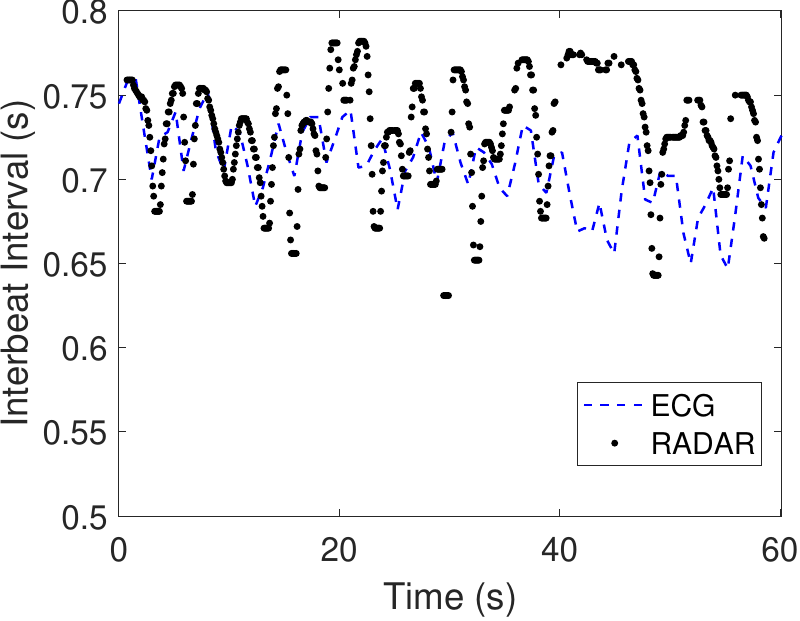}
        \subcaption{}
        \label{fig:C5_B2}
    \end{minipage}
    \begin{minipage}{0.49\linewidth}
        \centering
        \includegraphics[width = \linewidth]{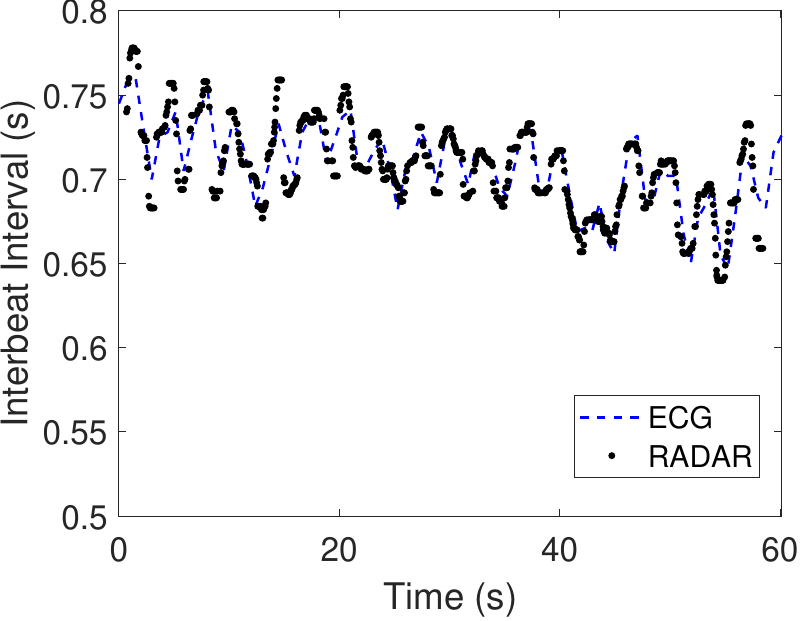}
        \subcaption{}
        \label{fig:P5_B2}
    \end{minipage}
    \caption{Participant A's IBI as estimated using (a) conventional method 3 and (b) proposed method 3.}
    \label{fig:CompareIBI_B2}
\end{figure}

\begin{figure}[tb]
    \centering
    \begin{minipage}{0.49\linewidth}
        \centering
        \includegraphics[width = \linewidth]{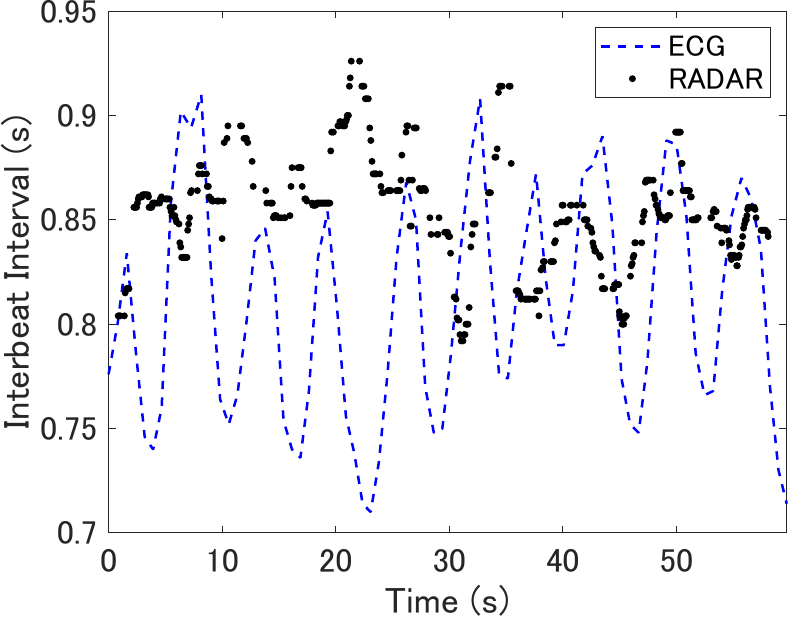}
        \subcaption{}
        \label{fig:C5_I1}
    \end{minipage}
    \begin{minipage}{0.49\linewidth}
        \centering
        \includegraphics[width = \linewidth]{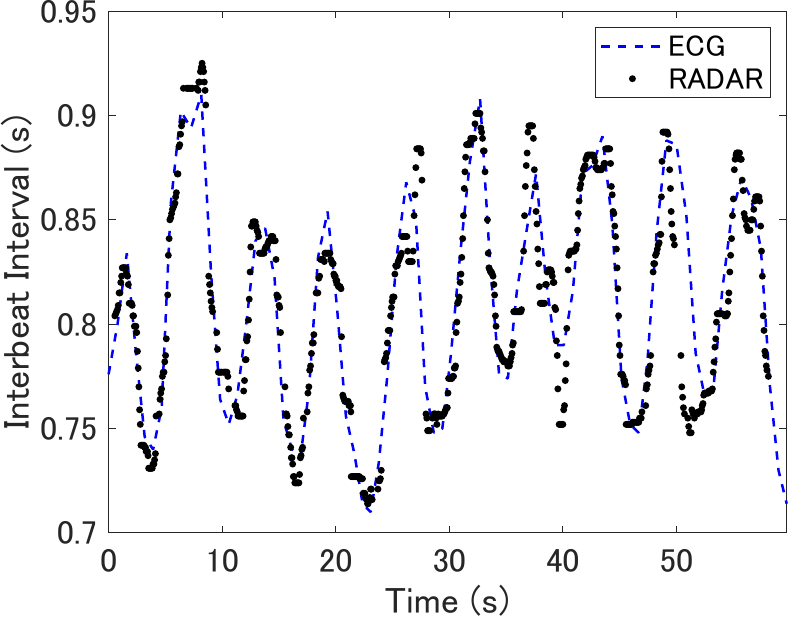}
        \subcaption{}
        \label{fig:P5_I1}
    \end{minipage}
    \caption{Participant B's IBI as estimated using (a) conventional method 3 and (b) proposed method 3.}
    \label{fig:CompareIBI_I1}
\end{figure}

As an example, Fig. \ref{fig:CompareIBI_B2} shows the IBIs that were estimated using both the conventional and proposed methods for participant A, where the CC, RMSE, and TCR values obtained were 0.15, 40.28 ms, and 77.50\% when using the conventional method, and 0.90, 11.76 ms, and 96.67\% when using the proposed method, respectively.
Similarly, Fig. \ref{fig:CompareIBI_I1} shows the IBIs that were estimated using both the conventional and proposed methods for participant B, where the CC, RMSE, and TCR values obtained were -0.29, 71.67 ms, and 51.26\% when using the conventional method, and 0.93, 19.46 ms, and 94.96\% when using the proposed method, respectively.
By comparing Fig. \ref{fig:C5_B2} with Fig. \ref{fig:P5_B2} and also comparing Fig.\ref{fig:C5_I1} with Fig. \ref{fig:P5_I1}, we can see that the proposed methods can estimate the IBIs accurately over the entire measurement period. This performance improvement over the conventional methods can be attributed to the enhanced heartbeat components obtained. 


\section{Conclusion}\label{section.5}
In this study, we have proposed a radar-based method for estimation of heart IBIs using a combination of the absolute value of the second derivative of the complex radar signal and the VME method.
Unlike many conventional methods, the proposed method does not estimate a body displacement waveform that is proportional to the phase of the complex radar signal. We have demonstrated that the heartbeat components were amplified effectively when using the second-derivative-based approach, which is emphasized selectively by the VME method, with a resulting improvement in accuracy when estimating the heart IBIs. By performing measurements on eleven participants using a 79 GHz array radar system, the proposed method was demonstrated to be able to estimate their heart IBIs using the topology method; this approach improved the average correlation coefficient of the actual and estimated IBIs by 0.2 and also reduced the average RMSE by 23\%. The proposed method is expected to make a major contribution to the development of a practical noncontact medical and healthcare monitoring system.

\section*{Acknowledgment}
\scriptsize
This work was supported in part by the SECOM Science and Technology Foundation, in part by the Japan Science and Technology Agency under Grants JPMJMI22J2 and JPMJMS2296, and in part by the Japan Society for the Promotion of Science KAKENHI under Grants 19H02155, 21H03427, 23H01420, 23K19119, 23K26115, and 24K17286.
The experimental protocol involving human participants was approved by the Ethics Committee of the Graduate School of Engineering, Kyoto University (permit nos. 201916, 202214, and 202223). Informed consent was obtained from all human participants in the study.
The authors thank Dr. Hirofumi Taki and Dr. Shigeaki Okumura of MaRI Co., Ltd. for providing technical advice.
The authors also thank all parties involved for their cooperation in conducting this research.
\normalsize

\balance

\end{document}